\documentclass[12pt,preprint]{aastex}
\usepackage{amsmath, amsthm, amssymb}
\usepackage{natbib}
\usepackage{lscape}

\begin{document}

\title{Solar cycle 25: another moderate cycle?}

\author{R.H. Cameron\altaffilmark{1}, J. Jiang\altaffilmark{2},
        M. Sch\"{u}ssler\altaffilmark{1}}
\altaffiltext{1}{Max-Planck-Institut f\"ur Sonnensystemforschung,
               Justus-von-Liebig-Weg 3, 37077 G\"ottingen, Germany}
\altaffiltext{2}{Key Laboratory of Solar Activity, National
Astronomical Observatories, Chinese Academy of Sciences, Beijing
100012, China}

\email{cameron@mps.mpg.de}

\begin{abstract}
{Surface flux transport simulations for the descending phase of cycle 24
using random sources (emerging bipolar magnetic regions) with
empirically determined scatter of their properties provide a prediction
of the axial dipole moment during the upcoming activity minimum together
with a realistic uncertainty range.  The expectation value for the
dipole moment around 2020 $(2.5\pm1.1\,$G) is comparable to that
observed at the end of cycle 23 (about $2\,$G). The empirical
correlation between the dipole moment during solar minimum and the
strength of the subsequent cycle thus suggests that cycle 25 will be of
moderate amplitude, not much higher than that of the current cycle.
However, the intrinsic uncertainty of such predictions resulting from
the random scatter of the source properties is considerable and
fundamentally limits the reliability with which such predictions can be
made before activity minimum is reached.}
\end{abstract}


\section{Introduction}
\label{sec:intro}

Owing to the growing practical importance of space weather, attempts to
predict future levels of solar activity have found much interest in
recent years. A considerable variety of methods has been suggested, of
which the so-called precursor methods proved most successful \citep[see
review by][]{Petrovay:2010}. These methods take the strength of the
magnetic field at the solar polar caps (or some proxy thereof, such as
the open heliospheric magnetic flux, the strength of the radial
interplanetary field, or the level of geomagnetic disturbances) during
activity minimum as indicator for the strength of the subsequent cycle.
The high empirical correlation between precursor level and cycle
strength \citep[e.g.][]{Wang:Sheeley:2009} has a theoretical basis in
the Babcock-Leighton scenario for the solar dynamo
\citep{Charbonneau:2010, Cameron:Schuessler:2015}. Thus a prediction of
the amplitude of the next activity maximum made at a time around
minimum, when the polar fields are fully developed, rests on rather
solid empirical and theoretical ground. The level of uncertainty of such
a prediction can be estimated from the data.

The question is whether and how a prediction at an earlier phase during
a cycle can be made. One possibility is to use surface flux transport
(SFT) simulations, which have quite successfully reproduced the observed
evolution of the large-scale magnetic field at the solar surface in the
course of the solar cycle, particulary also the evolution of the polar
fields and axial dipole moment \citep[e.g.][]{Wang:etal:1989,
Sheeley:2005, Mackay:Yeates:2012, Jiang:etal:2014a,
Upton:Hathaway:2014}. Such simulations also showed that randomness in
the properties of the magnetic flux sources in the form of emerging
bipolar magnetic regions (such as emergence latitude and tilt angle) has
a strong effect: single bipolar regions emerging near the equator can
significantly affect the level of the polar field (or axial dipole
moment) around solar minimum and thus influence the strength of the
subsequent cycle \citep{Jiang:etal:2014b, Jiang:etal:2015}.  Therefore,
any prediction of the polar field strength also needs to quantify the
uncertainty that arises from the random scatter of the source
properties.

In the work presented in this paper, we used empirically determined
statistical properties of emerging bipolar regions (sunspot groups)
during the declining phases of solar cycles and ran Monte-Carlo
ensembles of SFT simulations starting from observed synoptic maps of the
surface magnetic field. We thus obtained the axial dipole moment during
solar minima as an ensemble average over the realizations together with
an empirically based estimate of the uncertainty. We prefer to consider
the axial dipole moment and not the polar field strength since it is a
uniquely defined quantity and limits the effect of a hemispherically
asymmetric distribution of magnetic flux.  Around activity minima, the
axial dipole moment is dominated by the magnetic flux at the polar caps
and represents the open heliospheric flux.  We first tested our approach
by deriving postdictions for the axial dipole moment during the minima
of cycles 21-23, which can be compared with observations. We then ran
simulations for cycle 24 and obtained predictions and statistical
uncertainties for the dipole moment during the upcoming minimum around
2020.

\section{Surface flux transport (SFT) model}
\label{sec:SFT}

\subsection{SFT code}
\label{subsec:SFT_code}

The SFT code used is described in \citet{Baumann:etal:2004} and
\citet{Cameron:etal:2010}. It treats the evolution of the radial
component of the large-scale magnetic field at the solar surface
resulting from passive transport by convection, differential rotation,
and meridional flow. The corresponding magnetohydrodynamic induction
equation is given by
\begin{eqnarray}
\nonumber\frac{\partial B}{\partial t}=& &
-\Omega(\lambda)
          \frac{\partial B}{\partial \phi}
         - \frac{1}{R_\odot \cos\lambda}
              \frac{\partial}{\partial \lambda}[\upsilon(\lambda)
         B \cos \lambda] \\ \noalign{\vskip 2mm}
& & +\eta \left[\frac{1}{R_\odot^2 \cos{\lambda}}
                \frac{\partial}{\partial \lambda}\left(\cos\lambda
          \frac{\partial B}{\partial \lambda}\right) +
     \frac{1}{R_\odot^2 \cos^2{\lambda}}\frac{\partial^2 B}{\partial
     \phi^2}\right] + S(\lambda,\phi,t),
\label{eqn:SFT}
\end{eqnarray}
where $B(\lambda,\phi,t)$ is the radial component of the magnetic field.
$\lambda$ and $\phi$ are heliographic latitude and longitude,
respectively.  We used the profile of latitudinal differential rotation,
$\Omega(\lambda)$, determined by \citet{Snodgrass:1983} and took the
profile of the poleward meridional flow, $\upsilon(\lambda)$, from
\citet{Ballegooijen:1998}. The magnetic diffusivity
$\eta=250$~km$^2$s$^{-1}$ describes the random walk of the magnetic flux
elements as transported by supergranulation flows \citep{Leighton:1964}.
The source term, $S(\lambda,\phi,t)$, represents the emergence of
magnetic flux at the solar surface. The properties of the corresponding
bipolar regions were defined as described in \citet{Baumann:etal:2004}.
Further details are given in \citet{Cameron:etal:2010}.

\subsection{Source selection and initial condition}
\label{subsec:SFT_sources}

Our method requires us to provide synthetic source data (emerging
bipolar magnetic regions, BMRs) for the time period of the
prediction. The properties of these sources were chosen according to
their empirical statistics during the descending phases of
previous activity cycles.

The number of BMRs emerging per month of the simulation was taken to be
equal to $R/2.75$, where $R$ is the monthly sunspot number. This
calibration was carried out on the basis of the sunspot data for cycles
21--24 (between 1976 and 2016). We adopted the functional form
$R_G=f(t)+\Delta f(t)$, where $f(t)$ is given by Eq.~(1) of
\citet{Hathaway:etal:1994} and $\Delta f(t)$ represents the random
scatter of the time evolution.  We followed the procedure given in
\citet{Hathaway:etal:1994} to fix the independent parameters $a$
(amplitude) and $t_0$ (starting time) involved in $f(t)$ from
observational data for the cycles considered here. The obtained values
for $(a;t_0)$ are $(0.00351;1976.39)$ for cycle 21, $(0.00336;1986.16)$
for cycle 22, $(0.00229;1996.57)$ for cycle 23, and $(0.00103;2008.84)$
for cycle 24.  The standard deviation $\sigma_f(t)$ of $\Delta f(t)$ is
determined from the difference between the fit profiles $f(t)$ and the
observed profiles for cycles 21--23. It is well represented by
$\sigma_f(t) = 0.5 f(t)$.

For the latitudinal distribution of the sources as a function of cycle
phase we used the empirically derived functional forms for mean latitude
and width (standard deviation, $\sigma_\lambda$) of the activity belts
given by Eqs.~(4) and (5) of \citet{Jiang:etal:2011}. The area distribution
for the emerging BMRs and its dependence on cycle phase follows the
empirical profiles represented by Eqs.~(12)--(14) of
\citet{Jiang:etal:2011}.

The mean tilt angle, $\alpha_n(\lambda)$, of the emerging BMRs for cycle
number $n$ is assumed to follow Joy's law in the form
$\alpha_n(\lambda)=T_n\sqrt{|\lambda|}$. $T_n$ was taken from the linear
relation with maximum sunspot number given by Eq.~(15) of
\citet{Jiang:etal:2011}. For the scatter of the tilt angle,
$\sigma_\alpha$ (in degrees), we used the empirical relation with sunspot
umbral area, $A_U$, given by $\sigma_\alpha=-11\log A_U+35.$ as
determined by \citet{Jiang:etal:2014b}. To connect total sunspot area,
$A_S$, and BMR area with umbral area, we used the relation $A_S = 5 A_U$
\citep{Brandt:etal:1990}.

The relevant quantity for the contribution of a BMR to the axial
  dipole moment is the latitudinal separation of the two polarities,
  which depends on the tilt angle and on the distance between the
  polarities. For the latter we employ its empirical relation with the
  BMR area as derived by \citet{Cameron:etal:2010}. Here the BMR area is
  defined as the sum of sunspot area and plage area, using the
  relationship derived from observations by \citet{Chapman:etal:1997}.
  The distribution of latitudinal polarity distance derived by this
  procedure agrees well with that determined directly from historical
  sunspot data. This applies particularly to the tails of the
  distribution, which are most important for the buildup of the polar
  fields.

The procedure for defining the sources is then as follows: for each
month of the simulation, the number of emerging BMRs is determined
according to the value of $f(t)$ and a random deviation $\Delta
f(t)$. These BMRs are distributed randomly over the days of the
month. For each emerging BMR, its size is chosen randomly from the
(time-dependent) size distribution. Its latitude is given by the
(time-dependent) mean latitude plus a random component drawn from a
distribution with standard deviation $\sigma_\lambda$. The tilt angle
is chosen according to Joy's law plus a random component drawn from a
distribution with standard deviation $\sigma_\alpha$. The resulting
tilt angle is multiplied by a factor 0.7 to account for the effect of
latitudinal inflows towards active regions \citep{Cameron:etal:2010}.
Finally, the BMRs are distributed randomly over longitude and the two
hemispheres.

The introduction of the factor 0.7 is neccessary because the tilt
   angle distributions empirically determined by \citet{Jiang:etal:2014b}
   refer to the instant of maximum sunspot area during the evolution of
   an active region, which typically is a few days after the beginning
   of flux emergence. \citet[][see Fig.~8 therein]{Martin:Cameron:2016}
   have shown that the lateral inflows towards active regions
   significantly reduce the tilt angle after this moment.
   Quantitatively, the reduction is consistent with the factor 0.7
   assumed here, a value that also provided good reproduction of the
   observationally inferred polar fields (open heliospheric flux) during
   the activity minima between 1913 and 1986 with our flux transport
   model \citep{Cameron:etal:2010}.

Fig.~\ref{fig:emergence} shows an example for one realization of random
sources for cycle 24. The upper panel provides a comparison of the
actual monthly group sunspot number (black), one realization of the
random sources (red) and the average of 50 random realizations (blue).
The lower panel shows a butterfly diagram with the actual sunspot groups
(black crosses) and one realization of the random sources (red crosses).

We started the SFT simulations from observed synoptic magnetograms using
data from NSO/KPVT (available from 8/1976 until 8/2003), NSO/SOLIS
(8/2003 until present), SOHO/MDI (6/1996 until 11/2010), and SDO/HMI
(5/2010 until present).  All magnetogram data used were reduced to a
resolution of $1\deg$ in latitude and longitude, respectively.

\subsection{Error estimate}
\label{subsec:SFT_error}

For a sensible prediction, we need to quantify the uncertainty involved
in our extrapolation of the evolution of the axial dipole moment based
on SFT simulations with random sources. Two causes of error are
considered here: 1) uncertainty due to scatter in the properties of the
source BMRs (number, emergence location, size, and tilt angle) and 2)
uncertainty due to measurement error in the magnetogram used as initial
state. While (1) represents the intrinsic uncertainty resulting from the
random component of the solar dynamo process, the contribution of (2) is
due to imperfect measurement and depends on the instrument that provided
the data, the data analysis procedure, etc.  Since both contributions to
the total error are uncorrelated, they add quadratically.

The effect of the scatter in the source properties (1) was determined by
repeating each SFT simulation 50 times with different random
realizations of the sources. The uncertainty resulting from the error in
the initial magnetograms (2) was estimated by consideing the
averaged measured radial surface field,
\begin{equation}
  \overline{B_r}(t)=\frac{1}{4\pi} \int_{\pi/2}^{-\pi/2}\int_{0}^{2\pi}
   B_r(\lambda,t) \cos\lambda d\lambda d\phi,
\label{eq:net_flux}
\end{equation}
which should vanish since the magnetic field is divergence-free.  Any
deviation from zero must therefore be due to measurement error (e.g.,
instrumental noise, Zeeman saturation, $B$ angle correction).  We
estimate the rms error of $\overline{B_r}$ by considering a 19-month
running mean (indicated by angular brackets), $\varepsilon(t) =
\langle\overline{B_r}^2\rangle^{1/2}$.  Fig.~\ref{fig:rms} shows
$\varepsilon(t)$ for the various time series of synoptic magnetograms
considered here. As the instruments were improved over time and became
characterized better, $\varepsilon$ decreased since 1976 by about a
factor of 10 from values over $0.8\,$G to about $0.08\,$G. In the
estimate of the rms error we can safely neglect the average
$\langle\overline{B_r}\rangle$, which is small compared to $\varepsilon$.

In order to estimate the contribution of the measurement error in the
initial magnetogram to the uncertainty of the axial dipole moment
resulting from the SFT simulation, we need to consider the two
hemispheres separately. The relevant quantity is the rms of the {\em
error differences} between the hemispheres, $\varepsilon_{\rm
N-S}(t)$. Assuming that the errors on the two hemispheres are uncorrelated,
we have $\varepsilon_{\rm N-S}(t)=\varepsilon(t)$. If we assume further that
the statistical distribution of the error is uniform over the solar
surface, we can estimate the rms error of the axial dipole moment,
$\varepsilon_{\rm DM}(t)$, by running a SFT simulation without sources
and with an initial radial field equal to $\varepsilon(t_0)$ in the
northern and $-\varepsilon(t_0)$ in the southern hemisphere, where
$\varepsilon(t_0)$ is the rms error at the epoch of the initial
magnetogram.  This procedure draws upon the linearity of
Eq.~\ref{eqn:SFT}, which results in $\varepsilon_{\rm DM} \propto
\varepsilon_{\rm N-S} = \varepsilon$.

We note that there is possibly an additional contribution to the
intrinsic uncertainty.  \citet{Wang:etal:2009} proposed that
cycle-to-cycle variations of the mean meridional flow could be
responsible for the variability of the polar field amplitude. In the
absence of a sufficiently extended data base for the meridional flow, it
is difficult to judge the validity of this suggestion. If these
variations were of a random nature, they would (quadratically) add to
the uncertainty determined here.

\section{Results}
\label{sec:results}

\subsection{Axial dipole moment for cycles 21--23: postdiction}
\label{subsec:postdiction}

In order to test and validate our procedure, we carried out SFT
simulation runs with random sources for the descending phases of cycles
21, 22, and 23, for which we can directly compare the prediction
(actually: postdiction) from the simulation with the observed actual
evolution of the axial dipole moment. The simulations were started from
synoptic magnetograms obtained about 4 years prior to the subsequent
activity minima. The initial magnetograms correspond to Carrington
rotations CR1729 (Nov 25--Dec 22, 1982; NSO/KPVT) for cycle 21, CR1864
(Dec 24,1992--Jan 22, 1993; NSO/KPVT) for cycle 22, and CR2024 (Dec 5,
2004--Jan 3, 2005; SoHO/MDI) for cycle 23.  For each of the three
cycles, we carried out 50 SFT simulations with different realizations of
random sources. The results are shown in Fig.~\ref{fig:cyc21-23}. The
predictions for the axial dipole moment up to the corresponding solar
minimum are given by the averaged evolution for the 50 realizations each
(solid blue curves) in comparison with the actual data obtained from the
observed synoptic magnetograms. The uncertainty of the predictions was
determined according to the procedure outlined in
Sec.~\ref{subsec:SFT_error}. The total $\pm 2\sigma$-error is indicated by
blue shading. The dashed blue lines denote the $\pm 2\sigma$-error due to
source scatter alone, i.e., without the error in the initial
magnetograms.

The results show that the predictions agree with the actual observation
within the $\pm2\sigma$ uncertainty range. For cycle 22 though, this is
only marginally true owing to the early decay of the dipole moment
shortly after activity maximum.Such a deviation is expected to occur occasionally owing to the
   significant scatter in the tilt angle and other properties of
   BMRs. In particular, the axial dipole moment can be significantly
   affected by the emergence of single strongly tilted BMRs near the
   equator \citep{Cameron:etal:2013}.  In addition, fluctuations in the
   background meridional flow (not considered in our model) could also
   have contributed.

If a stochastic process indeed underlies the variability of the axial
dipole moment around solar minima, the prediction is expected to be
within the $\pm 2\sigma$ range for about 95\% of the cycles.
Considering only three cycles, of course we cannot make a definite
statement beyond noting that the comparison of our postdictions with the
observations is so far statistically consistent with the assumed
stochastic processes (basically source scatter).

\subsection{Dipole moment for cycle 24: predictor for cycle 25}
\label{subsec:prediction}

\begin{figure}
\begin{center}
\includegraphics[scale=0.8]{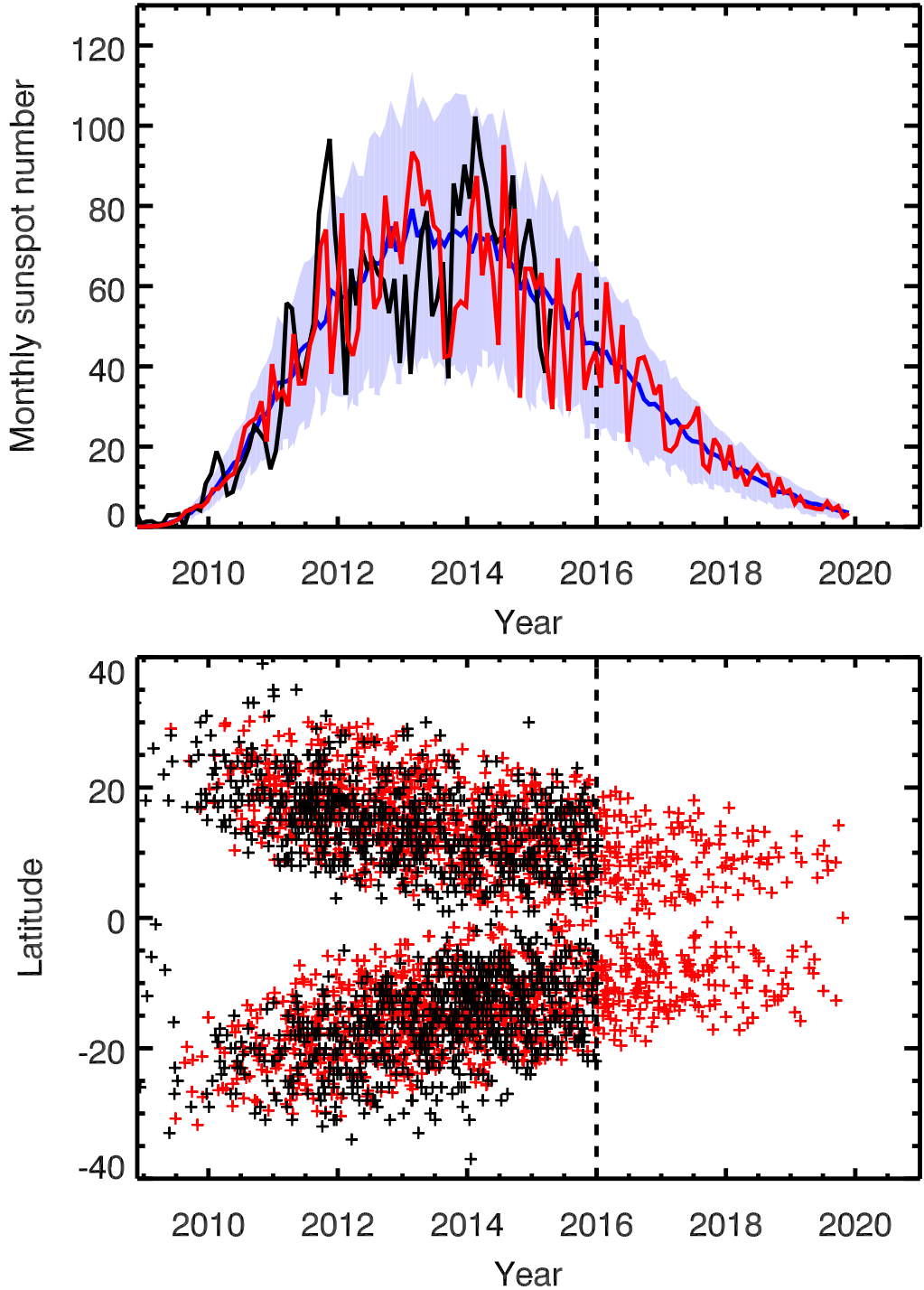}
\caption{Example of a realization of cycle 24 with random sources. {\em
         Upper panel:} monthly group sunspot number (black curve:
         observed; red curve: one random realization; blue curve:
         average of 50 random realizations). The shading indicates the
         $\pm 2\sigma$ variation corresponding to 50 random
         realizations. {\em Lower panel:} butterfly diagram of emerging
         sunspot groups (bipolar magnetic regions). Black crosses
         indicate observed sunspot groups; red crosses give one
         realisation of random sources, corresponding to the red curve
         in the upper panel.}
\label{fig:emergence}
\end{center}
\end{figure}

\begin{figure}
\begin{center}
\includegraphics[scale=0.8]{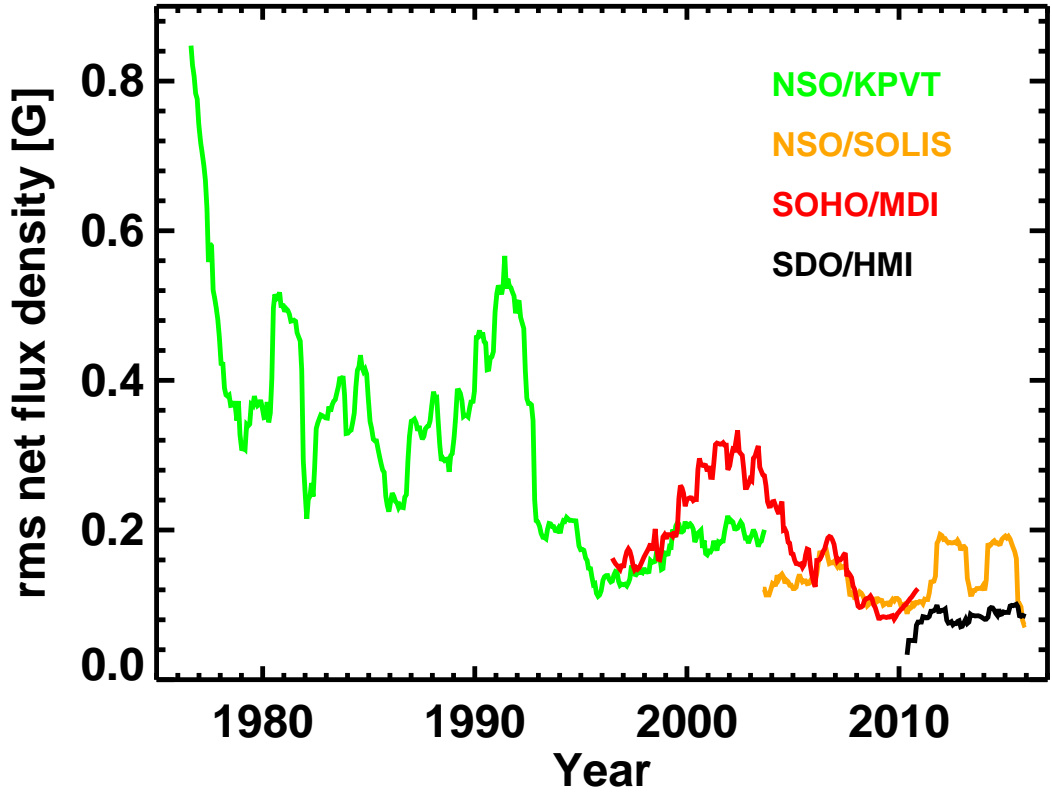}
\caption{RMS fluctuation of the surface-averaged radial magnetic field
         shown by synoptic magnetograms. The line colors indicate the
         data source as indicated in the legend.}
 \label{fig:rms}
\end{center}
\end{figure}

\begin{figure}
\begin{center}
\includegraphics[scale=0.8]{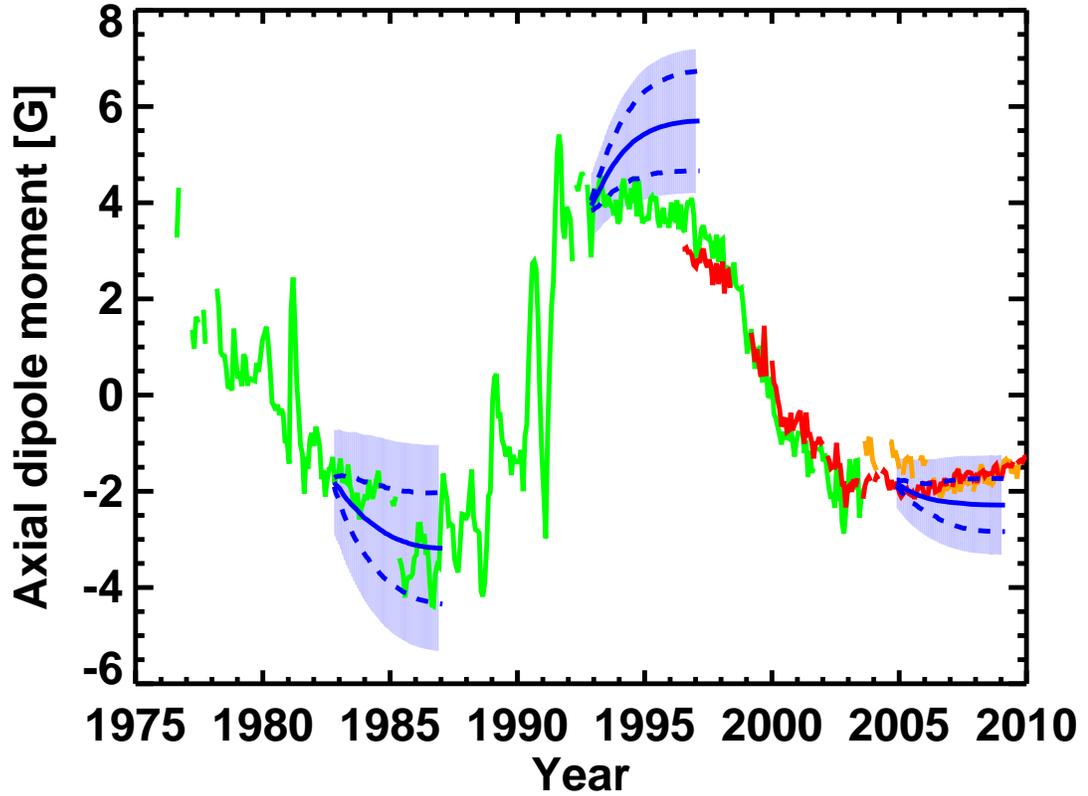}
\caption{Evolution of the axial dipole moment for cycles 21--23. Solid
  blue lines show the average of 50 SFT simulations with random
  sources. Blue shading indicates the total $2\sigma$ uncertainty
  range. The dashed blue lines goive the $2\sigma$ range for the
  intrinsic solar contribution (source scatter). The other colored lines
  correspond to the values determined from the various series of
  synoptic magnetograms (see legend in Fig.~\ref{fig:rms}).}
 \label{fig:cyc21-23}
\end{center}
\end{figure}

\begin{figure}
\begin{center}
\includegraphics[scale=0.8]{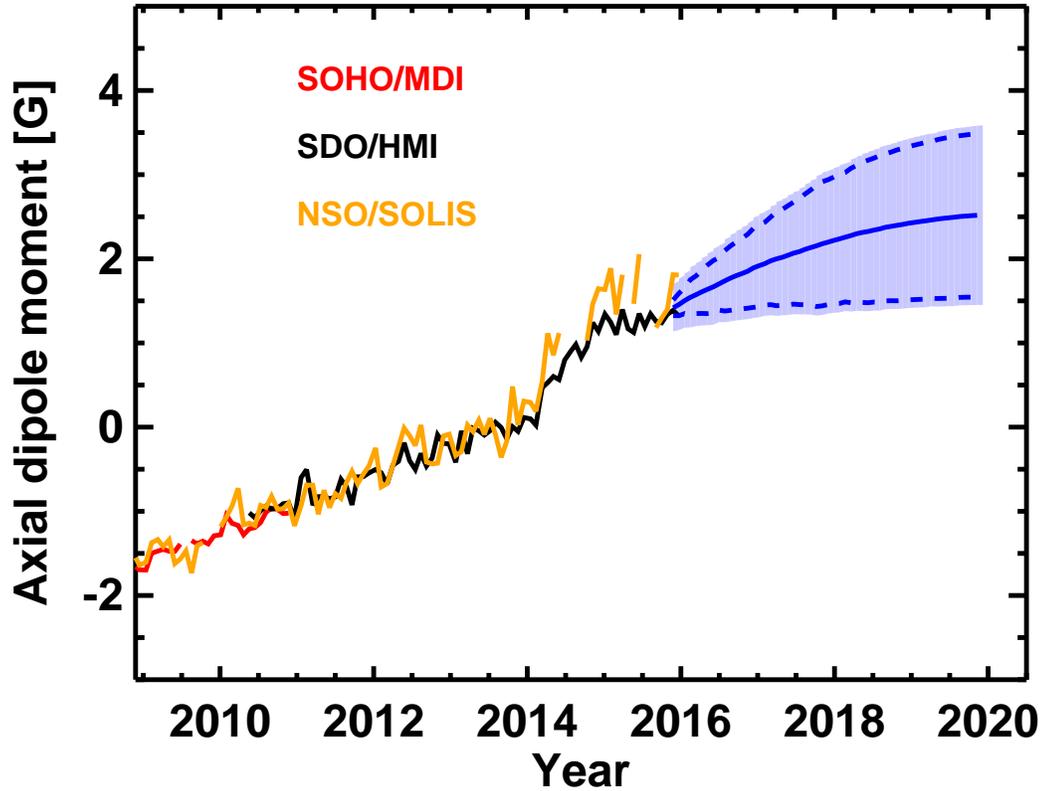}
\caption{Predicted evolution of the axial dipole moment until 2020
  (expected activity minimum of cycle 24) based on 50 SFT simulation
  with random sources starting from a synoptic magnetogram taken with
  SDO/HMI. The figure layout corresponds to that of
  Fig.~\ref{fig:cyc21-23}. Black and orange lines indicate values
  determined from SDO/HMI and NSO/SOLIS data, respectively.}
\label{fig:cyc24}
\end{center}
\end{figure}

We now consider the prediction for the axial dipole moment during the
minimum of the current cycle 24, expected to occur around the year 2020
(cf.~Fig.~\ref{fig:emergence}). As initial state for the SFT simulation
we used the synoptic magnetogram for Carrington rotation 2171 (Nov
28--Dec 25, 2015) from SDO/HMI. Using the same procedure as for the
previous cycles, we determined the predicted evolution of the axial
dipole moment, which is displayed in Fig.~\ref{fig:cyc24}.  During the
upcoming minimum around 2020, the expected value and $2\sigma$
uncertainty for the axial dipole moment is $2.5\pm1.1\,$G. The expected
value for the dipole moment is not much higher than that observed at the
end of cycle 23 ($\sim2\,$G) and significantly lower than those for
cycle 21 ($\sim3.5\,$G) and 22($\sim4.1\,$G). However, within the rather
high level of the $2\sigma$ uncertainty, the dipole moment could be as
high as that of cycle 21 or significantly lower than that of cycle 23.
Note also that (for reasons unknown to us) the NSO/SOLIS and SDO/HMI
data drifted apart by about $0.5\,$G. As a result, taking the NSO/SOLIS
magnetogram for CR2171 as initial state leads to a higher expectation
value of $3.1\pm1.1$G for the axial dipole moment around 2020.

Owing to the good (although not perfect) correlation between the open
heliospheric magnetic flux (strongly related to the axial dipole moment)
during solar minimum and the strength of the subsequent cycle
\citep{Wang:Sheeley:2009, Cameron:Schuessler:2012} we therefore expect
that cycle 25 will be of moderate strength, but possibly somewhat more
active than the current cycle. However, the uncertainty of this
prediction is considerable: within the $2\sigma$ range, cycle 25 could
also be as strong as cycle 22 or even considerable weaker than cycle 24.
This reflects the intrinsic limitation of such predictions resulting
from the random nature of flux emergence.

%\section{Conclusion}
%\label{sec:conclusion}

\begin{acknowledgements}
All authors contributed equally to the work presented in this paper.
SOHO is a project of international cooperation between ESA and NASA.
The SDO/HMI data are courtesy of NASA and the SDO/HMI team. The sunspot
records are courtesy of WDC-SILSO, Royal Observatory of Belgium,
Brussels. The National Solar Observatory (NSO)/Kitt Peak data used here
are obtained cooperatively by NSF-NOAO, NASA/Goddard Space Flight
Center, and the NOAA Space Environment Laboratory.  NSO/SOLIS data were
courtesy of NISP/NSO/AURA/NSF.  J. Jiang acknowledges the support by the
National Science Foundation of China (grants 11522325, 11173033) and by the
Youth Innovation Promotion Association CAS.
\end{acknowledgements}

%\bibliographystyle{apj}
%\bibliography{cyc25}

\begin{thebibliography}{24}
\expandafter\ifx\csname natexlab\endcsname\relax\def\natexlab#1{#1}\fi

\bibitem[{{Baumann} {et~al.}(2004){Baumann}, {Schmitt}, {Sch{\"u}ssler}, \&
  {Solanki}}]{Baumann:etal:2004}
{Baumann}, I., {Schmitt}, D., {Sch{\"u}ssler}, M., \& {Solanki}, S.~K. 2004,
  \aap, 426, 1075

\bibitem[{{Brandt} {et~al.}(1990){Brandt}, {Schmidt}, \&
  {Steinegger}}]{Brandt:etal:1990}
{Brandt}, P.~N., {Schmidt}, W., \& {Steinegger}, M. 1990, \solphys, 129, 191

\bibitem[{{Cameron} {et~al.}(2013){Cameron}, {Dasi-Espuig}, {Jiang}, {I{\c
  s}{\i}k}, {Schmitt}, \& {Sch{\"u}ssler}}]{Cameron:etal:2013}
{Cameron}, R.~H., {Dasi-Espuig}, M., {Jiang}, J., {I{\c s}{\i}k}, E.,
  {Schmitt}, D., \& {Sch{\"u}ssler}, M. 2013, \aap, 557, A141

\bibitem[{{Cameron} {et~al.}(2010){Cameron}, {Jiang}, {Schmitt}, \&
  {Sch{\"u}ssler}}]{Cameron:etal:2010}
{Cameron}, R.~H., {Jiang}, J., {Schmitt}, D., \& {Sch{\"u}ssler}, M. 2010,
  \apj, 719, 264

\bibitem[{{Cameron} \& {Sch{\"u}ssler}(2012)}]{Cameron:Schuessler:2012}
{Cameron}, R.~H. \& {Sch{\"u}ssler}, M. 2012, \aap, 548, A57

\bibitem[{{Cameron} \& {Sch{\"u}ssler}(2015)}]{Cameron:Schuessler:2015}
---. 2015, Science, 347, 1333

\bibitem[{{Chapman} {et~al.}(1997){Chapman}, {Cookson}, \&
  {Dobias}}]{Chapman:etal:1997}
{Chapman}, G.~A., {Cookson}, A.~M., \& {Dobias}, J.~J. 1997, \apj, 482, 541

\bibitem[{Charbonneau(2010)}]{Charbonneau:2010}
Charbonneau, P. 2010, Living Reviews in Solar Physics, 7, 3,
  http://www.livingreviews.org/lrsp

\bibitem[{{Hathaway} {et~al.}(1994){Hathaway}, {Wilson}, \&
  {Reichmann}}]{Hathaway:etal:1994}
{Hathaway}, D.~H., {Wilson}, R.~M., \& {Reichmann}, E.~J. 1994, \solphys, 151,
  177

\bibitem[{{Jiang} {et~al.}(2011){Jiang}, {Cameron}, {Schmitt}, \&
  {Sch{\"u}ssler}}]{Jiang:etal:2011}
{Jiang}, J., {Cameron}, R.~H., {Schmitt}, D., \& {Sch{\"u}ssler}, M. 2011,
  \aap, 528, A82

\bibitem[{{Jiang} {et~al.}(2014{\natexlab{a}}){Jiang}, {Cameron}, \&
  {Sch{\"u}ssler}}]{Jiang:etal:2014b}
{Jiang}, J., {Cameron}, R.~H., \& {Sch{\"u}ssler}, M. 2014{\natexlab{a}}, \apj,
  791, 5

\bibitem[{{Jiang} {et~al.}(2015){Jiang}, {Cameron}, \&
  {Sch{\"u}ssler}}]{Jiang:etal:2015}
---. 2015, \apjl, 808, L28

\bibitem[{{Jiang} {et~al.}(2014{\natexlab{b}}){Jiang}, {Hathaway}, {Cameron},
  {Solanki}, {Gizon}, \& {Upton}}]{Jiang:etal:2014a}
{Jiang}, J., {Hathaway}, D.~H., {Cameron}, R.~H., {Solanki}, S.~K., {Gizon},
  L., \& {Upton}, L. 2014{\natexlab{b}}, \ssr, 186, 491

\bibitem[{{Leighton}(1964)}]{Leighton:1964}
{Leighton}, R.~B. 1964, \apj, 140, 1547

\bibitem[{{Mackay} \& {Yeates}(2012)}]{Mackay:Yeates:2012}
{Mackay}, D. \& {Yeates}, A. 2012, Living Reviews in Solar Physics, 9, 6

\bibitem[{{Martin-Belda} \& {Cameron}(2016)}]{Martin:Cameron:2016}
{Martin-Belda}, D. \& {Cameron}, R.~H. 2016, \aap, 586, A73

\bibitem[{{Petrovay}(2010)}]{Petrovay:2010}
{Petrovay}, K. 2010, Living Reviews in Solar Physics, 7

\bibitem[{{Sheeley}(2005)}]{Sheeley:2005}
{Sheeley}, Jr., N.~R. 2005, Living Reviews in Solar Physics, 2, 5

\bibitem[{{Snodgrass}(1983)}]{Snodgrass:1983}
{Snodgrass}, H.~B. 1983, \apj, 270, 288

\bibitem[{{Upton} \& {Hathaway}(2014)}]{Upton:Hathaway:2014}
{Upton}, L. \& {Hathaway}, D.~H. 2014, \apj, 780, 5

\bibitem[{{van Ballegooijen} {et~al.}(1998){van Ballegooijen}, {Cartledge}, \&
  {Priest}}]{Ballegooijen:1998}
{van Ballegooijen}, A.~A., {Cartledge}, N.~P., \& {Priest}, E.~R. 1998, \apj,
  501, 866

\bibitem[{{Wang} {et~al.}(1989){Wang}, {Nash}, \& {Sheeley}}]{Wang:etal:1989}
{Wang}, Y.-M., {Nash}, A.~G., \& {Sheeley}, Jr., N.~R. 1989, Science, 245, 712

\bibitem[{{Wang} {et~al.}(2009){Wang}, {Robbrecht}, \&
  {Sheeley}}]{Wang:etal:2009}
{Wang}, Y.-M., {Robbrecht}, E., \& {Sheeley}, Jr., N.~R. 2009, \apj, 707, 1372

\bibitem[{{Wang} \& {Sheeley}(2009)}]{Wang:Sheeley:2009}
{Wang}, Y.-M. \& {Sheeley}, N.~R. 2009, \apjl, 694, L11

\end{thebibliography}

\end{document}